# Information Hiding as a Challenge for Malware Detection

**Wojciech Mazurczyk | Warsaw University of Technology**

**Luca Caviglione | National Research Council of Italy**

**Abstract.** Information hiding techniques are increasingly utilized by the current malware to hide its existence and communication attempts. In this paper we highlight this new trend by reviewing the most notable examples of malicious software that shows this capability.

## Introduction

We're experiencing an exponential growth in malicious software. According to the antivirus research firm AV-TEST, 2014 saw approximately 130 million new forms of malware, compared to just over 80 million in 2013 and about 30 million in 2012 (www.av-test.org/en/statistics/malware). Although the influx of malware has drawn the attention of security experts worldwide, the countermeasures that are currently available are progressively showing their limitations. For example, Symantec, one of the largest antivirus vendors, recently admitted that its products are able to detect only approximately 45 percent of new threats.[1] As a result, we should expect a relevant increase in the number of undiscovered types of malware.

Consider the case of the Regin Trojan, called a "top-tier espionage tool" by Symantec and other security companies. The sophistication of Regin and other malware such as Flame, Duqu, and Stuxnet leads industry experts to believe that they weren't created by "typical" cybercriminals for profit. Instead, they're thought to be created by nation-states to spy on a wide range of international targets and eventually launch attacks if necessary. Regin has been used since at least 2008 to spy on several international targets including government and business organizations, infrastructure operators, researchers, and private individuals. Its six-year period of hidden activity raises the question: How can malware developers avoid detection for long periods of time?

## Information Hiding

Providing a clear answer is difficult, but the most common arguments consider the increasing degree of sophistication of new threats, such as modular design to enable customization (seen in Regin, Flamer, and Weevil) or multistage loading architectures in which each stage is hidden and encrypted (seen in Regin, Stuxnet, and Duqu). In this article, we highlight the importance of understanding information-hiding techniques in malware. These techniques have often been neglected by the security community, but are widely used to exfiltrate data and make security threats stealthier by postponing their detection for as long as possible.

*Information hiding* is part of a wide spectrum of methods that are used to make data difficult to notice. This practice shouldn't be confused with encryption, in which the content is unreadable, as it is instead overt. Such mechanisms are often used jointly to ensure that a conversation remains unreadable. *Steganography* is one of the most well-known subfields of information hiding and aims to cloak secret data in a suitable carrier. Historical examples include the use of tattoos or invisible ink to hide a conversation from unauthorized observers.[2]

Typically, to exchange secrets, the involved parties must agree on a preshared scheme and embed the secrets in a carrier: the greater the carrier's popularity, the better its masking capacity. Too many alterations would reveal the presence of hidden information, thus limiting the amount of data that can be covertly transmitted. For example, using too many least significant bits of an image's pixels as the carrier can reveal the secret data due to visible artifacts. Onkar Dabeer and his colleagues' "Detection of Hiding in the Least Significant Bit" shows a representative method of detecting this scheme.[3]

Networks play an important role in modern malware, making network steganography a crucial tool: in this case, the secret is injected into network traffic. For example, the data can be cloaked by manipulating the content of unused flags within headers or by modulating the inter-packet time (IPT) of network flow datagrams. In the latter case, a sender can encode bits of information in previously agreed-upon IPT values. Similar to the LSB modification of image's pixels example, overly aggressive deviations would make it possible to differentiate the hiding process from normal jitter events. Therefore, hidden channels are typically characterized by a low bandwidth, often ranging from a few to a few hundreds bits per second.

Today, many other methods enable covert communications among desktops or digital devices, including generating inaudible sounds or utilizing a smartphone's sensors to receive a sequence that activates a threat. Malware can use information-hiding techniques to cloak its existence, making it harder to detect. Having a better understanding of these types of malware will help security professionals detect, mitigate, and prevent attacks.

**Roots of the New Trend**

Advancements in security systems over the past 15 years have forced malware developers to investigate new possibilities to make their "products" stealthier. Although it's difficult to determine the origin of information-hiding techniques, the first massive usage of these techniques can be traced back to 2006, when Operation Shady RAT led to attacks against numerous institutions worldwide and inflicted damage for months.[4] Years later, security experts agreed that the main program responsible for this attack was the phishing virus Trojan.Downbot.[5] This virus created a back door and then downloaded files appearing as real HTML pages or JPEG images. These files were encoded with commands that would allow remote servers to gain access to local files on the infected host computer.

Other notable examples of information hiding–capable malware include Regin and Linux.Fokirtor, which use network traffic to covertly leak data, and Alureon, Duqu, Lurk, and Trojan.Zbot, which use digital images as hidden data carriers. Even when rudimentary, new threats exploiting some form of information hiding continue to be discovered, as seen in Soundcomber and AirHopper, which modify the status of shared hardware/software resources to exfiltrate confidential data, and in Feederbot, W32.Morto, and Smuggler, which manipulate network traffic for this purpose.

Smartphones are better suited than desktops to exploit information hiding because they natively incorporate cameras, GPS, WLAN, Bluetooth, cellular networks, and other various sensors.[6] Even when using legacy general packet radio service or connectivity with bandwidth scarcity—in which case, data leaking could be very slow or impracticable—the availability of different carriers could provide an effective workaround. Furthermore, after their success with desktops, malware developers turned a significant portion of their attention to mobile devices, leading to a 1,800 percent increase in mobile malware over the past two years, as reported by McAfee.[7] Threats using information hiding on mobile platforms could be the next great challenge for security researchers.

**Information-Hiding Malware: A Classification**

Nearly all information hiding–capable malware was discovered between 2011 and 2014, with a peak in 2014. Table 1 shows the most popular types and proof-of-concept implementations proposed by the research community. However, we consider only examples that are sufficiently mature to be deployed in real scenarios. A convenient way to organize existing hiding-capable malware is according to the methodology used to implement covert communications. As such, we introduce three major groups:

- group 1—methods that hide information by modulating the status of shared hardware/software resources,

- group 2—methods that inject secret data into network traffic, and

- group 3—methods that embed secret data by modifying a digital file's structure or using digital media steganography, for example, by manipulating image pixels or sound samples.

Groups 2 and 3 contain techniques that are primarily used to increase the stealthiness of communications carrying commands or leaked data that are mainly observed in malware-targeting desktops. Group 1 includes mechanisms that bypass a security perimeter, such as a sandbox, or enable communications from or to an isolated source or destination, for example, two disconnected devices located on the same workbench. In this case, the prime targets are smartphones and mobile devices.

Next, we describe what we consider the three most meaningful examples for each group.

*Table 1. The most popular and recent information hiding–capable malware. Real-life malware means real information hiding-capable malware found by antivirus vendors in the Internet; academic means proof-of-concept malware proposed by academic community.*

| *Malware name or developers* | *Group* | *Discovery/proposal date* | *Desktop (D) or mobile (M)* | *Real-life malware (R) or academic (A)* |
|---|---|---|---|---|
| Soundcomber | 1, 2 | Feb. 2011 | M | A |
| Trojan.Downbot | 3 | May 2011 | D | R |
| Feederbot | 2 | Aug. 2011 | D | R |
| W32.Morto | 2 | Aug. 2011 | D | R |
| Alureon | 3 | Sept. 2011 | D | R |
| Duqu | 3 | Sept. 2011 | D | R |
| Gasior and Yang[14,15] | 2 | Oct. 2011/Dec. 2012 | M | A |
| Trojan:Android/FakeRegSMS.B | 3 | Jan. 2012 | M | R |
| Marforio and his colleagues[16] | 1 | Dec. 2012 | M | A |
| Sensor-based malware | 1 | May 2013 | M | A |
| KINS Trojan (variant of Zeus) | 3 | June 2013 | D | R |
| Linux.Fokirtor | 2 | Sept. 2013 | D | R |
| Lalande and Wendzel[17] | 1 | Sept. 2013 | M | A |
| Inaudible sound-based malware | 1 | Nov. 2013/Aug. 2014 | D/M | A |
| Lurk | 3 | Feb. 2014 | D | R |
| Trojan.Zbot | 3 | Mar. 2014 | D | R |
| Oldboot.B | 3 | Apr. 2014 | M | R |
| AirHopper | 1 | Oct. 2014 | D/M | A |
| Smuggler[18] | 2 | Nov. 2014 | D/M | A |
| Multilayer .NET malware | 3 | Nov. 2014 | D | R |
| Regin | 2 | Nov. 2014 | D | R |

**Group 1**

Researchers' increasing attention combined with Android's open source nature has allowed the development of many instances of proof-of-concept information hiding–capable mobile malware. A prime example is Soundcomber,[8] which covertly transmits the buttons pressed during a call, for example, when entering a PIN for a bank service. Notably, it uses information hiding to bypass the security framework of mobile OSs. In fact, the malware could have insufficient privileges to access the network to exfiltrate data, so it can use a "colluding" application to leak data outside the device.

Soundcomber utilizes several information-hiding methods to form four local covert channels whose range is limited to the single device. The covert techniques exploit the most popular smartphone functionalities such as vibration or volume settings (one process differentiates vibration or volume status, and another infers secret data bits from this event), screen state (secret bits are transferred by acquiring and releasing the wake-lock permission that controls the screen state), and file locks (secret data are exchanged between the processes by competing for a file lock).

As we mentioned, another relevant field in which information hiding can be used is the covert transmission of data from and to devices that are physically isolated from other peers. For instance, Luke Deshotels uses standard smartphone speakers to transmit data via ultrasonic sounds.[9] This technique can cover distances up to 30 meters with a rate of 9 bits per second. Similarly, AirHopper enables infected devices to communicate by modulating the graphics processing unit load to emit electromagnetic signals.[10] In this case, the coverage is reduced to 7 meters, but the rate is in the range of 100 to 500 bits per second.

Finally, in "Sensing-Enabled Channels for Hard-to-Detect Command and Control of Mobile Devices," Ragib Hasan and his colleagues demonstrate a method to trigger attacks on a large population of infected smartphones in the same geographic area.[11] Latent malware could be activated by using built-in sensors listening to ad hoc hidden stimuli, such as a song with a particular pattern, vibrations from a subwoofer, or the ambient light from a TV or a monitor.

**Group 2**

In 2011, Symantec announced the discovery of the worm W32.Morto, which propagates using a vulnerability in the remote desktop protocol. To communicate with command and control (C&C), it uses domain name system (DNS) records. Specifically, W32.Morto exploits the TXT record, which was originally introduced to contain text readable by humans. W32.Morto queries for a DNS TXT record (not for a domain to IP lookup), and then validates and decrypts the returned data. The obtained information typically yields a binary signature and an IP address where the worm can retrieve another malware to execute.

The recently identified Linux.Fokirtor is a Trojan virus that opens a back door and allows attackers to remotely compromise a host. Symantec reported that the malware was utilized in

May 2013 to attack one of the largest hosting providers and focused on stealing confidential customer information such as credentials and emails.[12] As cybercriminals realized that their target network was generally well protected, they hid malware communications in an innocent secure shell and other server process network traffic. In addition to this information-hiding technique, Linux.Fokirtor used the Blowfish encryption algorithm to cipher stolen data or other communications with its C&C server.

In November 2014, Regin took malware stealthiness a step further. It utilized many sophisticated mechanisms, including antiforensics capabilities, a custom-built encrypted virtual file system, and an alternative encryption (RC5 variant). It also exploited information hiding in network traffic to covertly communicate with its C&C server by tunneling secrets in Internet Control Message Protocol/ping traffic and embedding commands in HTTP cookies or in custom TCP segments and UDP datagrams.

**Group 3**

In the second half of 2011, the Laboratory of Cryptography and System Security in Budapest, Hungary, discovered malware generating strange files with the prefix "~DQ"; as a result, it was named Duqu.[13] It bears many resemblances to the famous Stuxnet worm, which was likely developed to attack Iran's nuclear infrastructure. Duqu is generally considered the precursor to a future Stuxnet-like attack. Duqu's main aim is to gather information about industrial control systems. To exfiltrate secrets, it encrypted data, which was appended at the end of innocent digital images and then sent over the Internet to a C&C server. This approach postponed the worm's detection because the images containing leaked information were hidden in the bulk of actual digital pictures. In the same period, a variant of Alureon used a comparable technique.

In February 2014, malware called Lurk was found spreading via websites using <iframes> or an Adobe Flash exploit. A thorough analysis revealed the use of steganography to embed encrypted URLs in an image by manipulating pixels. Such information is then used to retrieve an additional payload.

Another approach uses a variant of the Trojan.Zbot malware, which was first detected in 2014. This version downloaded innocent-looking JPEG images depicting sunsets or cats that contained a list of IP addresses to be inspected, mainly pointing at financial institutions. Once users visit any of the listed destinations, the malware proceeds to steal their confidential information, such as access credentials.

Regin, Duqu, and Lurk are real-life examples of what security experts should expect to see daily in the future. In fact, even if the information-hiding methods utilized in current and future malware aren't yet very sophisticated, they could become dangerous in the next few years if state-of-the-art academic solutions are considered.

**Future Trends**

Due to the rich availability of options and the masking features offered by the massive utilization of the Internet, malware developers could find the highest potential in network steganography. Although early steganographic techniques focused only on modifying unused fields of TCP/IP headers (such as the type of service field of IPv4, which is rarely set by routers), more recent and sophisticated methods include, but aren't limited to, the exploitation of flows produced by popular services such as Skype and BitTorrent. Furthermore, network traffic produced by popular online games can be used to covertly exchange data, even in devices with limited capabilities, such as gaming consoles. In fact, the signaling used to locate players in a first-person shooter game can be an effective carrier. To this end, some bits of the set of coordinates and angles can be used to hide data.[3]

In addition, because smartphones are complete computing platforms, they can leverage all the techniques presented in this article and combine them with a rich set of sensors, offering essentially unlimited options for covertly communicating with the surrounding environment. From this perspective, we can envision the following future trends:

- New information-hiding techniques will be continually introduced, and their degree of sophistication will increase. Hence, future malware-related traffic could be harder to detect.

- Information hiding offers a decoupled design. Therefore, it can be easily incorporated into every type of malware to provide stealthy communication of both control commands and the exfiltration of confidential user data as well as communication from isolated environments or networks.

- Information hiding–capable malware can remain cloaked for a long period of time while slowly but continuously leaking sensitive user data. Thus, this type of malware must be considered a new advanced persistent threat and must be addressed properly.

**A** long-term solution to these trends is to consider the potential vulnerabilities enabling covert communications from the very early design phases of desktop and mobile platforms, services, and protocols. For existing devices, especially smartphones, a short-term approach would require some form of ad hoc mitigation, at least for the most hazardous threats. However, this a posteriori approach is very difficult because there aren't yet any universal countermeasures. We hope that raising awareness and understanding of these information-hiding techniques will help researchers and security experts develop the necessary countermeasures.

***Wojciech Mazurczyk*** *is an associate professor at the Institute of Telecommunications in the Faculty of Electronics and Information Technology at the Warsaw University of Technology. He's also an associate technical editor for* IEEE Communications Magazine*. Contact him at wmazurczyk@tele.pw.edu.pl.*

***Luca Caviglione*** *is a researcher at the National Research Council of Italy. He holds several patents and is an associate editor of Wiley's* Transactions on Emerging Communications Technologies*. Contact him at luca.caviglione@ge.issia.cnr.it.*